\documentclass[aps,pre,twocolumn,showpacs,10pt,superscriptaddress]{revtex4-1}

\usepackage{graphicx}
\usepackage{dcolumn}
\usepackage{bm}
\usepackage{amsmath}
\usepackage{amssymb}
\usepackage{float}
\usepackage[titletoc,title]{appendix}
\usepackage{multirow}

\newcolumntype{P}[1]{>{\centering\arraybackslash}p{#1}}

\begin{document}


\author{Jirawat Tangpanitanon}
\email{cqtjt@nus.edu.sg}
\affiliation{Centre for Quantum Technologies, National University of Singapore, 3 Science Drive 2, Singapore 117543}
\author{Supanut Thanasilp}
\affiliation{Centre for Quantum Technologies, National University of Singapore, 3 Science Drive 2, Singapore 117543}
\author{Ninnat Dangniam}
\affiliation{
Department of Physics and Center for Field Theory and Particle Physics, Fudan University, Shanghai 200433, 
China}
\author{Marc-Antoine Lemonde}
\affiliation{Centre for Quantum Technologies, National University of Singapore, 3 Science Drive 2, Singapore 117543}
\affiliation{State Key Laboratory of Surface Physics, Fudan University, Shanghai 200433, China}
\author{Dimitris G. Angelakis}
\email{dimitris.angelakis@nus.edu.sg}
\affiliation{Centre for Quantum Technologies, National University of Singapore, 3 Science Drive 2, Singapore 117543}
\affiliation{School of Electrical and Computer Engineering, Technical University of Crete, Chania, Greece 73100}


\title{Expressibility and trainability of parameterized analog quantum systems for machine learning applications}

\date{\today}

\begin{abstract}
Parameterized quantum evolution is the main ingredient in variational quantum algorithms for near-term quantum devices. In digital quantum computing, it has been shown that random parameterized quantum circuits are able to ‘express’ complex distributions intractable by a classical computer, leading to the demonstration of quantum supremacy. However, their chaotic nature makes parameter optimization challenging  in variational approaches. Evidence of similar classically-intractable expressibility has been recently demonstrated in analog quantum computing with driven many-body systems. A thorough investigation of trainability of such analog systems is yet to be performed. In this work, we investigate how the interplay between external driving and disorder in the system dictates the trainability and expressibility of interacting quantum systems. We show that if the system thermalizes, the training fails at the expense of the a large expressibility, while the opposite happens when the system enters the many-body localized (MBL) phase. From this observation, we devise a protocol using quenched MBL dynamics which allows accurate trainability while keeping the overall dynamics in the quantum supremacy regime. Our work shows the fundamental connection between quantum many-body physics and its application in machine learning. We conclude our work with an example application in generative modeling employing a well studied analog many-body model of a driven Ising spin chain. Our approach can be implemented with a variety of available quantum platforms including cold ions, atoms and superconducting circuits

\end{abstract}

\maketitle

\section{Introduction}

The recent achievement of quantum supremacy \cite{2019_martinis_nat}, the ability of quantum systems to compute tasks that are intractable by a classical computer, stands as an important milestone for noisy intermediate-scale quantum (NISQ) devices \cite{ 2018_preskill_quantum}. A common approach to operate NISQ devices is to implement variational quantum algorithms (VQAs), where a classical feedback loop is used to passively correct the noise in the quantum device \cite{Benedetti_2019,Moll_2018, McClean_2016_njp}. VQAs have been implemented to tackle a wide range of problems, from quantum chemistry \cite{2020arXiv200404174A,2018_Hempel_PRX,2017_ibm_nat,2017_Li_PRX, 2016_Omalley_PRX,2014_jeremy_natcom}, machine learning \cite{Zhueaaw9918, 2019_ibm}, quadratic binary optimization \cite{2019arXiv190602700P, 2020arXiv200404197A, 2017_Otterbach_ArXiV},  to high energy physics \cite{2019_zoller}. 

One of the key questions for NISQ devices is whether they can provide provable quantum advantage for real-world problems. A hint to answer this question lies in the ability of NISQ devices to efficiently explore Hilbert space. For example, in quantum chemistry, NISQ devices can produce highly-entangled variational ansatzes, such as unitary coupled clusters, that cannot be efficiently represented on a classical computer \cite{2017arXiv170102691R}. In machine learning, quantum circuits have been proven to have more `expressive power' than any classical neuron networks \cite{PhysRevResearch.1.033063, 2018_tao_arxiv,2019_kashfi_arxiv}. This means that those circuits can produce complex probability distributions that cannot be efficiently sampled from a classical computer. 

Similar to classical variational algorithms, VQAs rely on `good' ansatzes that can efficiently capture the answer of a given problem. In the case when such ansatz is not known or implementable, it is  desirable to exploit high expressibility of some NISQ devices to generate an unbiased guess. The latter is known as `hardware efficient' \cite{2017_ibm_nat}. A common feature of this approach is to explore the chaotic dynamics which allows the system to quickly explore the entire Hilbert space. However, this chaoticity also makes it difficult, if not impossible, to classically optimize the system since it is highly sensitive to any small changes in the parameters. In the digital case, hardware efficient ansatzes suffer from the barren plateaus problem \cite{2018_hartmut_natcom, 2020arXiv200100550C}, where the landscape of the cost function becomes exponentially flat as the number of qubits increases. Hence, finding the right VQA for a given problem is an emerging art of balancing expressibility, implementability and trainability of the NISQ devices. 

Analog quantum simulators stand out from their digital counterpart when it comes to implementability~\cite{2012_zoller_natphy, 2012_lewenstein_rpp,2014_dieter_epj}. Here, a quantum device is built to mimic a specific Hamiltonian, which requires significantly less control than universal quantum circuits. State-of-the-art quantum simulators have already been able to produce dynamics intractable by existing classical algorithms~\cite{2016_Choi_Sci}. Quantum supremacy in analog simulators have also been proven in 2D Ising lattice \cite{2018_eisert_prx,PhysRevLett.118.040502}, cluster states \cite{2019_patron_arxiv}, and more recently in periodically-driven quantum many-body systems \cite{2020arXiv200211946T}. Hybrid analog-digital approaches for VQAs have been explored in Refs \cite{2019arXiv190602700P, 2019_ibm, 2018_Hempel_PRX,2017_ibm_nat,2017_Li_PRX,PhysRevA.101.022305, PhysRevResearch.2.013012, 2019arXiv191209331G}.

In this work, we analyze the expressibility and trainability of analog quantum devices focusing on parameterized driven quantum many-body systems. We show that these properties are intimately related to phases of the system. We focus on four generic phases depending on whether the dynamics is thermalized or many-body localized (MBL) \cite{RevModPhys.91.021001, 2015_huse_arcmp} and whether a continuous drive is applied. As an example, we consider the standard Ising chain, globally driven by an external magnetic field. We find that, evolving under the dynamics resulting from a series of quenches between randomized disorder configurations, the system in all four phases are capable of reaching the quantum supremacy regime, illustrating its high expressibility beyond a classical computer. We then devise a simple sequential training protocol to train the system for generative modeling tasks in machine learning. We show that the chaoticity in the thermalized phase prevents the training as in the digital case. However, the integrability of the MBL within each quench increases drastically the trainability of the system. The final learning accuracy depends solely on the phase of the system. 

\begin{table*}
\begin{tabular}{p{6cm} P{2cm} P{2cm} p{0.5cm} P{2cm} P{2cm}}
\hline
\multirow{2}{*}{Features}&\multicolumn{2}{c}{$f(t)=0$}&\multirow{2}{*}{} & \multicolumn{2}{c}{$f(t)\neq 0$}\\
\cline{2-3} \cline{5-6} 
& Thermalized & MBL &  & Thermalized & MBL\\
\hline 
Statistics of $\hat{H}_{\rm ave}(\underline{\theta}_m)$& GOE & POI  & & - &- \\
Statistics of $\hat{U}(\underline{\theta}_m)$& - & -  &  & COE  & POI \\
Statistics of $\hat{\mathcal{U}}(\underline{\Theta}_M)$ with $M\gg 1$& CUE & CUE & &  CUE &CUE \\ \\

High expressibility (quantum supremacy) & yes & yes  &  &  yes  &  yes\\ 
Trainability for generative modeling & no & yes &  &  no  & yes (best) \\ 
\hline
\end{tabular}
\caption{A summary of statistics, expressibility, and trainability in the four regimes, defined by whether $\hat{U}(\underline{\theta}_m)$ is thermalized or MBL and whether $f(t)=0$ or $f(t)\neq 0$. The symbol `-' indicates that the statistics is not defined.}
\label{tb:summary}
\end{table*}

\section{DRIVEN ANALOG QUANTUM SYSTEMS AND THEIR STATISTICS}

In this section, we study the many-body dynamics of generic parameterized quantum systems and the different statistics associated with their phases. We then analyze a specific example of driven quantum Ising chain which will be used for the analysis of the expressibility and trainability in the following sections.

\subsubsection{General framework}

We consider fully general quenched quantum many-body systems $|\psi(\underline{\Theta}_M)\rangle=\hat{\mathcal{U}}(\underline{\Theta}_M)|\psi_0\rangle$, where $|\psi_0\rangle$ is an initial product state, $\underline{\Theta}_M$ is a vector containing all variational parameters during the evolution and $M$ is the number of times the system is quenched. The unitary time evolution is  
\begin{equation}
	\hat{\mathcal{U}}(\underline{\Theta}_M)=\hat{U}(\underline{\theta}_{M})\hat{U}(\underline{\theta}_{M-1})...\hat{U}(\underline{\theta}_1),
\end{equation}
where $\underline{\Theta}_M=\{\underline{\theta}_m\}_{m=1}^M$ and each quench/layer is obtained from a time-dependent Hamiltonian $\hat{H}(\underline{\theta}_m,t)$, i.e.
\begin{equation}
	\hat{U}(\underline{\theta}_m)=\hat{\mathcal{T}}\exp\left(-i\int_0^T\hat{H}(\underline{\theta}_m,t)dt\right),
\end{equation}
with $m\in\{1,2,...,M\}$, $\hat{\mathcal{T}}$ being the time-ordering operator and $T$ being the evolution time during each layer. The Hamiltonian is further decomposed as 
\begin{equation}
	\hat{H}(\underline{\theta}_m,t)=\hat{H}_{0}(\underline{\theta}_m)+f(t)\hat{V},
\end{equation}
where $\hat{H}_{0}(\underline{\theta}_m)$ is a static Hamiltonian, $\hat{V}$ is the driving Hamiltonian such that $\left[\hat{H}_{0}(\underline{\theta}_m),\hat{V}\right]\neq 0$. The modulation $f(t)$ is an oscillating function with the period $T$. We require that the time-averaged Hamiltonian $\hat{H}_{\rm ave}(\underline{\theta}_m)=\frac{1}{T}\int_0^T\hat{H}(\underline{\theta}_m,t)dt$ is many-body \cite{eisert}. 

\subsubsection{The four phases of $\hat{U}(\underline{\theta}_m)$}

 In the following, we will refer to $\hat{U}(\underline{\theta}_m)$ as `thermalized' if any observations made on $|\psi'_{M'}\rangle=\hat{U}(\underline{\theta}_m)^{M'}|\psi_0\rangle$ with $M'\to \infty$ can be obtained from the micro-canonical ensemble predictions associated with the energy $\bar{E}(\underline{\theta}_m)\pm \Delta E$, where $\bar{E}(\underline{\theta}_m)=\langle \psi_0 |\hat{H}_{\rm eff}(\underline{\theta}_m)|\psi_0\rangle$. The effective Hamiltonian $\hat{H}_{\rm eff}(\underline{\theta}_m)$ is defined such that $\hat{U}(\underline{\theta}_m)\equiv\exp\left[-i\hat{H}_{\rm eff}(\underline{\theta}_m)T\right]$. Most quantum many-body systems follow this property according to the eigenstate thermalization hypothesis (ETH) \cite{2008_maxim}. Likewise, we refer to $\hat{U}(\underline{\theta}_m)$ as `many-body localized' if the above is not true due to large disorder \cite{2016_Alessio_AiP}. Partial experimental signatures of the MBL have been observed in cold neutral atoms \cite{Schreiber842,PhysRevLett.116.140401,Lukin256,2017_bloch_natphy, 2016_Choi_Sci}, superconducting circuits \cite{2017_Roushan_Sci,PhysRevLett.120.050507}, and trapped ions \cite{2016_monroe_nat}. 
 
 We can now define the four regimes or `phases' of $\hat{U}(\underline{\theta}_m)$ in the above sense according to whether the dynamics is thermalized or MBL and whether $f(t)$ is zero or non-zero. To allow non-trivial dynamics within each layer, we require $2\pi/T$ to be smaller than a typical energy gap of $\hat{H}_{\rm ave}(\underline{\theta}_m)$. We assume that all $\hat{U}(\underline{\theta}_m)$'s in $\hat{\mathcal{U}}(\underline{\Theta}_M)$ belong to the same phase for simplicity.  
 
 Let us explore the various statistics associated with the four phases, starting with the $f(t)= 0$ case in which $\hat{H}_{\rm eff}(\underline{\theta}_m)=\hat{H}_{\rm ave}(\underline{\theta}_m)$. For the thermalized dynamics, the statistics of $\hat{H}_{\rm ave}(\underline{\theta}_m)$ follows the Gaussian orthogonal ensemble (GOE) \cite{2016_rigol_ap}. This is the ensemble of matrices whose entries are independent normal random variables subjected to the orthogonality constraint. This randomness is a signature of quantum chaos, which is a crucial ingredient for thermalization \cite{2016_Alessio_AiP}. A large disorder can prevent the system from thermalization leading to MBL dynamics. In this case, the eigenenergies of $\hat{H}_{\rm ave}(\underline{\theta}_m)$ follow the Poison (POI) statistics, indicating that they are uncorrelated. 
 
 In the driven case, i.e. $f(t)\neq 0$, the statistics are defined at the level of the unitary operator $\hat{U}(\underline{\theta}_m)$, as it is generally not possible to have access to $\hat{H}_{\rm eff}$. For the driven thermalized dynamics, the statistics of $\hat{U}(\underline{\theta}_m)$ follows the circular orthogonal ensemble (COE) \cite{2014_Rigol_PRX}. This is the ensemble of matrices whose entries are independent complex normal random variables subjected to the orthogonality and the unitary constraints. Unlike the GOE, the COE is intimately related to the infinite-temperature ensemble and is not possible to obtain without a drive \cite{2014_Rigol_PRX}. As before, a large disorder can prevent thermalization even with $f(t)\neq 0$, leading to the POI statistics of the quasi-energies (to be defined later) \cite{2015_Ponte_PRL,2016_Abanin_AoP}. A summary of all the statistics is given in Table \ref{tb:summary}.

\subsubsection{Driven disordered quantum Ising chains} 

To illustrate the four generic phases, we will work on a specific example of driven quantum Ising chains with
\begin{align}
	\hat{H}_{0}(\underline{\theta}_m)&=\sum_{i=1}^L\theta_{i,m}\hat{Z}_i+J\sum_{i=1}^{L-1}\hat{Z}_i\hat{Z}_{i+1}+\frac{h}{2}\sum_{i=1}^L\hat{X}_i\\
	\hat{V} &= \sum_{i=1}^L\hat{X}_i,
\end{align}
where $f(t) = -\frac{F}{2}\cos(\omega t)$, $\omega=2\pi/T$, $\underline{\theta}_m=\{\theta_{i,m}\}_{i=1}^L$, $L$ is the number of spins, $\{\hat{X}_i,\hat{Z}_i\}$ are Pauli's operators acting on site $i$, $J$ is the interaction strength, $h$ is a static magnetic field and $F$ is the driving amplitude. The parameters $\{\theta_{i,m}\}$ are `varied' by randomly drawing them from a uniform distribution in the range $\left[0,W\right]$ where $W$ is the disorder strength. This allows us to vary the parameters without changing the phase of the system. The dimension of the Hilbert space is $N=2^L$. The initial state $|\psi_0\rangle$ is prepared as a product state where each spin points along the $+z$ direction. This simple model has been implemented in various quantum platforms, including Rydberg atoms \cite{2017_lukin_nat}, trapped ions \cite{2017_monroe_nat} and superconducting circuits \cite{2017_Otterbach_ArXiV}. 

The standard way to analyze the statistics of the system is to define the level statistics $\Pr(r_{\alpha})$ as the normalized distribution of 
\begin{equation}
r_{\alpha}\equiv\frac{\min(\Delta_{\alpha+1},\Delta_{\alpha})}{\rm max (\Delta_{\alpha+1},\Delta_{\alpha})}, 
\end{equation}
where $\Delta_{\alpha}=\mathcal{E}_{\alpha+1}-\mathcal{E}_{\alpha}$ is the level spacing with $\mathcal{E}_{\alpha+1}>\mathcal{E}_{\alpha}$ and $\alpha=1,2,..,2^L-1$. In the $f(t)=0$ case, $\{\mathcal{E}_{\alpha}\}$ are eigenenergies of $\hat{H}_{\rm ave}(\underline{\theta}_m)$. In the $f(t)\neq 0$ case, $\{\mathcal{E}_{\alpha}\}$ are quasi-energies, defined such that $\{\exp\left(-i\mathcal{E}_{\alpha}T\right)\}$ are eigenvalues of $\hat{U}(\underline{\theta}_m)$. Not only that $H_{\rm eff}\neq H_{\rm ave}$ in the driven case, but the quasi-energies are also defined in the limited range $\mathcal{E}_\alpha \in \left[0,2\pi\right)$. This ‘energy folding’ has profound impact on the resulting statistic.

\begin{figure}
\includegraphics[width=1.0\columnwidth]{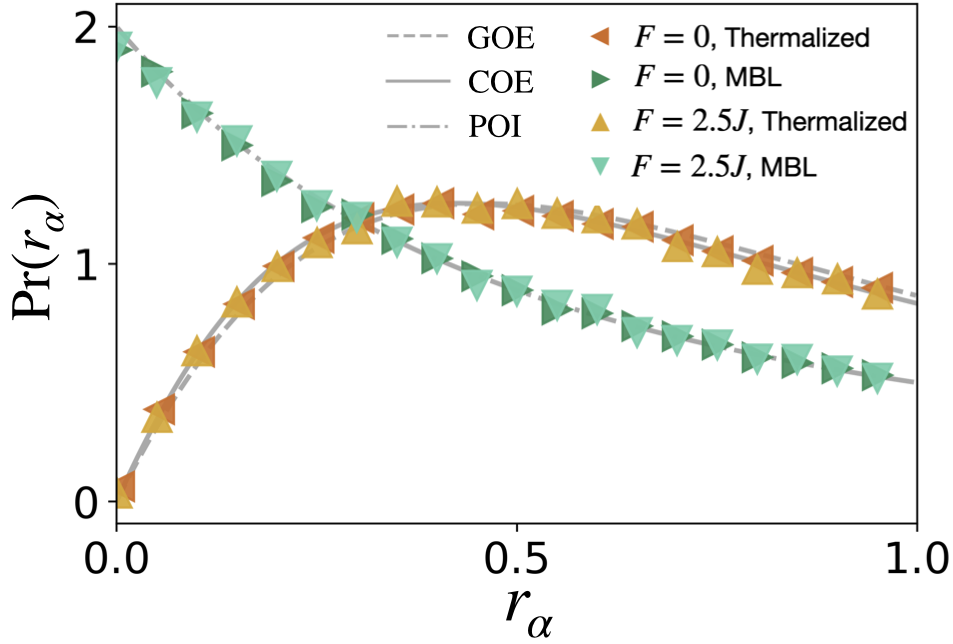}
\caption{Level statistics of $\hat{H}_{\rm ave}(\underline{\theta}_m)$ for $f(t)=0$ and $\hat{U}(\underline{\theta}_m)$ for $F=2.5J$. The thermalized and the MBL phases are obtained with $W=1J$ and $W=20J$,  respectively. ($L=9$, $\omega=8J$, $h=2.5J$, 500 disorder realizations).}
\label{fig2}
\end{figure}

In Fig. \ref{fig2}, we show the level statistics for $F=0$ and $F=2.5J$ with $W=1J$ and $W=20J$. For a small disorder $W=1J$, the level statistics of $\hat{H}_{\rm ave}(\underline{\theta}_m)$ and $\hat{U}(\underline{\theta}_m)$ agree with the predictions from the GOE and the COE, respectively. For a large disorder $W=20J$, the level statistics of both $\hat{H}_{\rm ave}(\underline{\theta}_m)$ and $\hat{U}(\underline{\theta}_m)$ follows the POI distribution, as expected.  

\section{Expressibility of driven quantum-many body systems}

In this section, we show that, given a large number of quenches $M$, the overall dynamics described by $\hat{\mathcal{U}}(\underline{\Theta}_M)$ for all four phases is capable of reaching the quantum supremacy regime, implying high expressibility of our system beyond a classical computer.

\subsubsection{Expressibility and quantum supremacy}

Expressibility is the term used in machine learning to describe the range of the resulting functions that a model can compute \cite{2016arXiv160605336R}. In the context of quantum computing, expressibility relates to how much a quantum system can explore the Hilbert space \cite{2019arXiv190510876S}. For example, product state ansatz have a lower expressibility than tensor-network ansatz, due to their inability to capture entangled states \cite{2019arXiv190703741G}. 

The concept of quantum supremacy and expressibility are interconnected. In random quantum circuit proposals for quantum supremacy, universal set of quantum gates are designed such that the system is chaotic and quickly explores the entire Hilbert space over time \cite{2017_Harrow_Nat}. Consequently, it is impossible for a classical computer to efficiently reproduce its output distribution, unless the polynomial hierarchy collapses. Hence, random quantum circuits with $L \gtrsim 100$ qubits have higher expressibility than any possible models, implementable on a classical computer. 

Let us consider the task of approximating $p(\underline{z};\underline{\Theta}_M)$ up to additive error, i.e.
\begin{equation}
	\sum_{\underline{z}\in\{0,1\}^L}|p(\underline{z};\underline{\Theta}_M)-q(\underline{z})|\leq \nu,
\end{equation}
where $\nu$ is a positive constant, $\{\underline{z}\}$ are output bit-strings measured in the computational basis, $p(\underline{z};\underline{\Theta}_M)=|\langle \underline{z}|\psi(\underline{\Theta}_M)\rangle|^2$ is the exact output probability, and $q(\underline{z})$ is the approximated value obtained from a classical / quantum device. In principle, a quantum device can satisfy this condition by directly implementing $\hat{\mathcal{U}}(\underline{\Theta}_M)$ in the hardware and measure the output multiple times to construct $q(\underline{z})$. To show that a classical computer cannot do the same efficiently unless the polynomial hierarchy collapses, one need to show that (i) it is $\#$P-hard to approximate $p(\underline{z};\underline{\Theta}_M)$ up to multiplicative error \cite{doi:10.1137/0214060}, i.e.
\begin{equation}
	 |p(\underline{z};\underline{\Theta}_M)-q(\underline{z})|\leq \eta p(\underline{z};\underline{\Theta}_M)
\end{equation}
for some $\eta$ and (ii) the output probability anti-concentrates \cite{Hangleiter2018anticoncentration}, i.e.
\begin{equation}
	{\rm Pr} \left( p(\underline{z};\underline{\Theta}_M) > \frac{\delta}{N} \right) \geq \gamma,
\end{equation}
where $\delta$, $\gamma$ are some constants. We refer interested readers to Ref. \cite{2018_eisert_prx, haferkamp2019closing, bouland2019complexity} for the derivation of how these two conditions lead to the proof of quantum supremacy.

\subsubsection{Achieving quantum supremacy with quenched quantum many-body systems}

The $\#$P-hardness to approximate $p(\underline{z};\underline{\Theta}_M)$ up to multiplicative error has been shown (for the worse instance) in the case where it results from a unitary evolution that follows the circular unitary ensemble (CUE) statistics \cite{bouland2019complexity, 2018_hartmut_natphy}. The CUE is the ensemble of matrices whose entries are independent complex normal random variables subject to the unitary constraint \cite{2010_haake}. Such statistics can be probed from both the previously defined level statistics ${\rm Pr}(r_{\alpha})$ and the distribution ${\rm Pr}(c=|\langle \underline{z}|\mathcal{E}_{\alpha}\rangle|^2)$ of the eigenstates $|\mathcal{E}_{\alpha}\rangle$ of $\hat{\mathcal{U}}(\underline{\Theta}_M)$. 

\begin{figure*}
\includegraphics[width=18cm,height=8.cm]{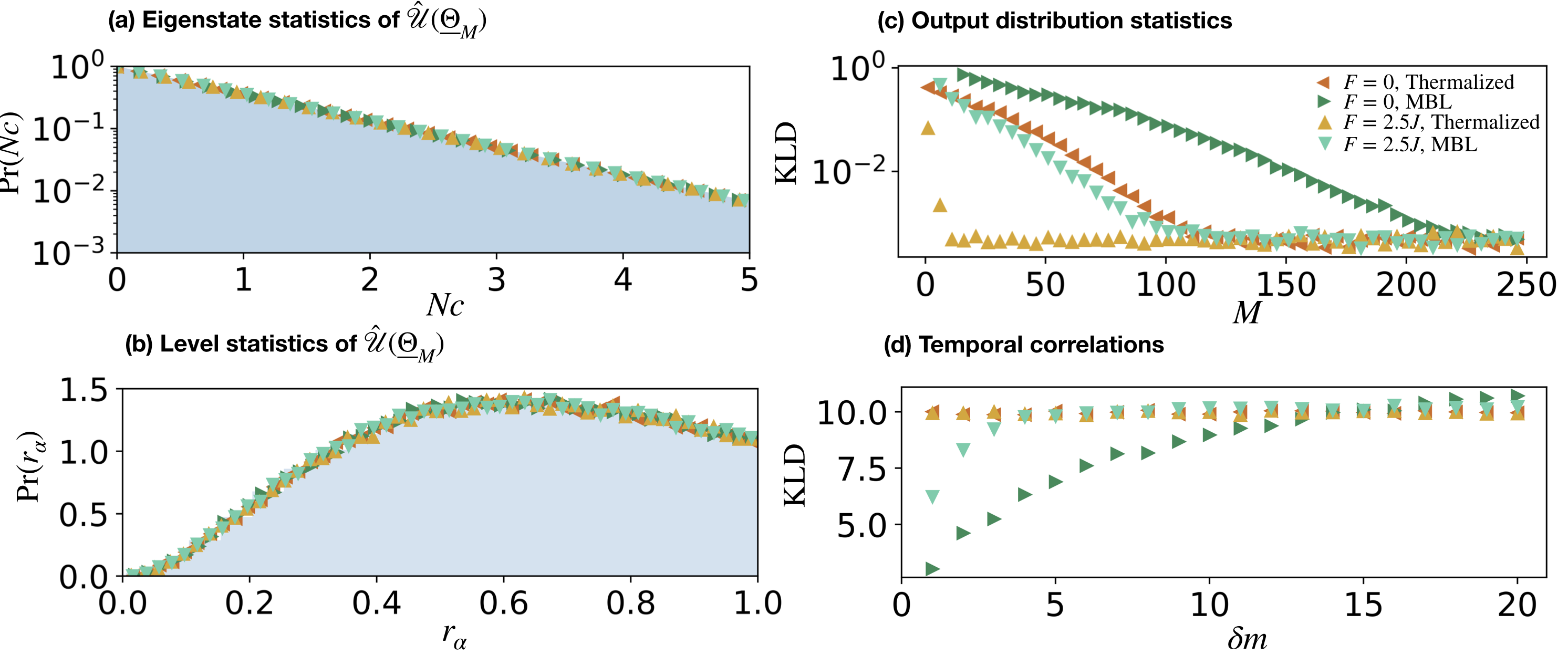}
\caption{\textbf{Statistics of parameterized analog quantum many-body evolution: }(a) and (b) shows the eigenstate distribution ${\rm Pr}(Nc)$ and the level statistics ${\rm Pr}(r_{\alpha})$ for the four phases of $\hat{U}(\underline{\theta}_m)$, respectively with $M=400$. The shaded areas are the predictions from the CUE statistics. (c) The KLD of the output distribution from the Porter-Thomas distribution as a function of $M$. (d) The KLD of $p(\underline{z};\underline{\Theta}_{m+\delta m})$ from $p(\underline{z};\underline{\Theta}_{m})$ as a function  of $\delta m$. The KLD is averaged over $M\in\left[378,400\right)$ for a given $\delta m$. The thermalized and and the MBL phases are obtained with $W=1J$ and $W=20J$,  respectively. ($L=9$, $\omega=8J$, $h=2.5J$, 500 disorder realizations).}
\label{fig3}
\end{figure*}

Fig. \ref{fig3}(a) and (b) show the statistics of the eigenstates and the quasi-energies of $\hat{\mathcal{U}}(\underline{\Theta}_M)$ in the four regimes at $M=400$, respectively. It can be seen that in all cases the results match with the CUE statistics, indicating the $\#$P-hardness to approximate the resulting $p(\underline{z};\underline{\Theta}_M)$ up to multiplicative error. Our finding agrees with Ref \cite{PhysRevA.97.023604}, which shows that random quenches in atomic Hubbard and spin models with long-range interactions lead to the $n$-design property. The $n$-design ensemble produces the CUE when $n\to \infty$ which happens in the long-time limit \cite{Harrow2009}.

In Fig. \ref{fig3}(c), we plot the Kullback-Leibler (KL) divergence of the output distribution ${\rm Pr}(p)$ from the Porter-Thomas distribution, ${\rm Pr}_{\rm PT}(p)=Ne^{-Np}$. The latter implies that the system explores the entire Hilbert space. (Here, we drop the argument $\underline{\Theta}_M$ for brevity). The Porter-Thomas distribution satistifies the anti-concentration condition since ${\rm Pr}_{\rm PT}\left(p > \frac{1}{N}\right) = \int_{Np=1}^{\infty} d(Np) e^{-Np} = 1/e$ \cite{2018_hartmut_natphy}. From Fig. \ref{fig3}(c), it can be seen that the system in all four phases reaches the Porter-Thomas distribution over time with different timescales. The thermalized case with $F=2.5J$ reaches it first at $M\sim 10$. The thermalized case with $F=0$ and the MBL case with $F=2.5J$ have a similar convergence rate and saturate at $M\sim 100$. The MBL with $F=0$ has the slowest rate and saturates at $M\sim 250$. This is expected as MBL dynamics localizes the system, while the drive $F$ `heats up' the system leading to de-localization. 

Fig. \ref{fig3}(a)-(c) provides evidences that $|\psi(\underline{\Theta}_M)\rangle$ cannot be efficiently approximated by a classical computer. This suggests that, for a large number of qubits, our system in all phases have higher expressibility than any classical models.  

\section{Trainability of driven analog quantum-many body systems}

In the context of machine learning, having a model with large expressibility is necessary but not sufficient as the model also need to be trainable. We here address the interplay between expressibility and trainability for the four generic phases of driven analog many-body systems discussed so far. Interestingly, we show that the external drive and the temporal correlations between different quenches in the MBL phase are the key ingredients to combine those two crucial characteristics.

\subsubsection{Generative modeling in classical machine learning}

\begin{figure}
\includegraphics[width=1\columnwidth]{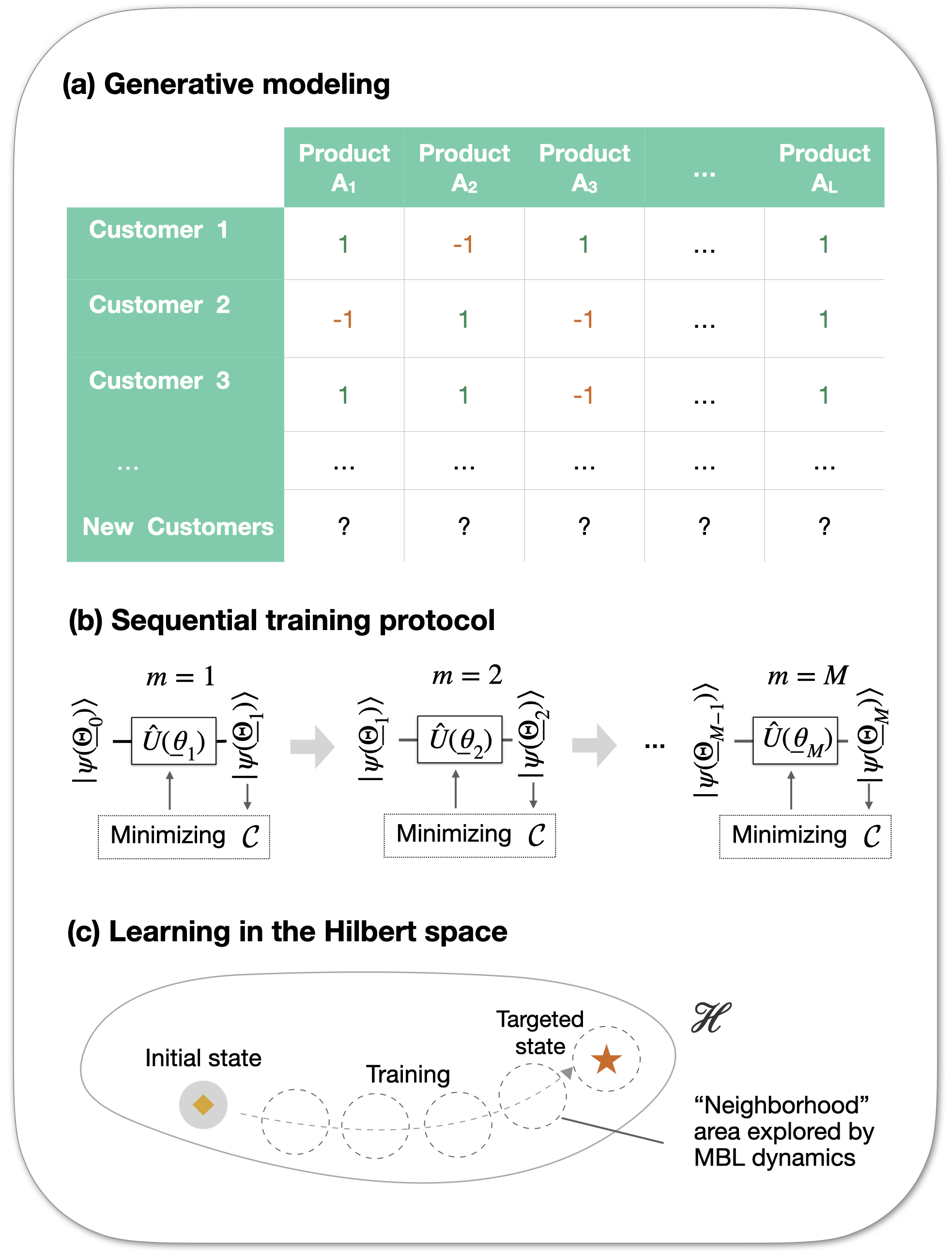}
\caption{\textbf{Machine learning with a driven analog quantum processor:} (a) A table demonstrating a real-world application of generative modeling tasks in machine learning. Each customer is asked to rate whether he/she likes ($+1$) or dislikes ($-1$) a given product. (b) A sketch of optimization loops used in the training protocol. (c) A diagram showing the movement of the system in the Hilbert space during the training in the MBL phase. }
\label{fig4}
\end{figure}

As a testbed to analyse the trainability of our model, we solve a generative modeling problem in machine learning \cite{2020_alpaydin}. The latter is an unsupervised task, meaning that the training data are unlabelled. The goal is to find the unknown probability distribution, $Q(\underline{z})$, underlying the training data. Here, the data is a set of binary vectors $\{\underline{z}\}_{\rm data}=\{\underline{z}_1,\underline{z}_2,...\}$. For example, it can represent the opinions of a group of customers on a set of $L$ different products, as depicted in Fig. \ref{fig4}(a). The opinion of the customer $i$ is represented by a binary vector $\underline{z}_i=\left[z_{i1},z_{i2},...,z_{iL}\right]$ where $z_{ij}=1$ if he/she likes the product $j$ and $-1$ otherwise. After knowing $Q(\underline{z})$, the company can generate new data from this distribution and recommends products with $+1$ score to new customers. 

In this section we use an artificial dataset as a working example. To assure the generality of the data, we assume that $Q(\underline{z})$ is the Boltzmann distribution of classical Ising spins with all-to-all connectivity, \textit{i.e.},
\begin{equation}
Q(\underline{z}) = \frac{1}{Z}e^{- E(\underline{z}) / k_B T_0}, 
\label{eq:q}
\end{equation}
where $Z=\sum_{\underline{z}} \exp{(- E(\underline{z}) / k_B T_0)}$ is the partition function, $k_B$ is the Boltzmann constant, $T_0$ plays the role of a temperature, and  
\begin{equation}
E(\underline{z}) = \sum_{i=1}^La_i z_i + \sum_{\langle i,j\rangle} b_{ij}z_iz_{j}
\label{eq:bm}
\end{equation}
with $a_i$, $b_{ij}$ being random numbers between $\pm J/2$. This model is known as the Boltzmann machine which is one of the standard types of artificial neuron networks used in machine learning and has been shown to capture a wide range of real-world data \cite{ACKLEY1987522}. Its quantum version has been studied in \cite{PhysRevX.8.021050, 2019_yudong}.

\subsubsection{Sequential training scheme using an analog quantum model}

Classically, the distribution of $\{\underline{z}\}_{\rm data}$ can be obtained by first guessing a model $P_{\rm model}(\underline{z};\underline{\Theta})$, such as the Poisson or the Boltzmann distribution, which has some variational parameters $\underline{\Theta}$. The `training' is done by minimizing the cost function, which is the KL divergence of $P_{\rm model}(\underline{z};\underline{\Theta})$ from $\tilde{Q}(\underline{z})$ using either gradient descent or gradient-free optimization algorithms. Here, $\tilde{Q}$ is the normalized histogram of $\{\underline{z}\}_{\rm data}$.

In our case, we show how the distribution of $\{\underline{z}\}_{\rm data}$ can be recovered as the output probability $p(\underline{z};\underline{\Theta}_M)$ of the driven quantum Ising chain. This approach is also known as the Born's machine \cite{2019_kashfi_arxiv}. Our goal here is to guide or ‘train’ the quantum system to a specific point in the Hilbert space such that $p(\underline{z};\underline{\Theta}_M)=Q(\underline{z})$. Our training protocol, depicted in Fig. \ref{fig4}(b), takes place as follows:
\begin{enumerate}
\item Initialize the system at $|\psi(\underline{\Theta}_{m})\rangle=|\psi_0\rangle$ with $m=0$ and $\underline{\Theta}_0=\{\}$.
\item Evolve the system by one layer $|\psi(\underline{\Theta}_{m+1})\rangle=\hat{U}(\underline{\theta}_{m+1})|\psi(\underline{\Theta}_m)\rangle$ with $\underline{\Theta}_{m+1}=\{\underline{\theta}_{m+1}\}\cup \underline{\Theta}_m$, and then measure $p(\underline{z};\underline{\Theta}_{m+1})$ to compute $\mathcal{C}$.
\item Repeat the step (2) $D$ times with different disorder realization $\underline{\theta}_{m+1}$. In the thermalized case, the system will randomly explore the entire Hilbert space in this step. However, in the MBL case, the system will only explore the Hilbert space locally near $|\psi(\underline{\Theta}_m)\rangle$ allowing systematic optimization, see Fig. \ref{fig4}(c).   
\item Choose the disorder realization in the step (3) that minimizes $\mathcal{C}$, then update $m\to m+1$. This will `move' the state in the most promising direction in the Hilbert space. 
\item Repeat the step (3)-(4) until convergence.
\end{enumerate}

We note here three characteristics of our training protocol. First, it is sequential since not all parameters in $\underline{\Theta}$ are updated at the same time, making them easier for classical optimization. Second, although the parameters are randomly drawn during the training, our optimization is done systematically in the Hilbert space. This makes an important difference to the usual optimization approaches which are done in the parameter space \cite{2018_hartmut_natcom, PhysRevX.8.021050}. Third, a large fraction of  results is `thrown away' in the step (3). Although in principle this data can be utilized to improve the training efficiency, it is our goal to keep the training protocol as simple as possible, so that the focus is made on distinct learning behaviors displayed by each phase.

\begin{figure*}
\includegraphics[width=16cm,height=7.cm]{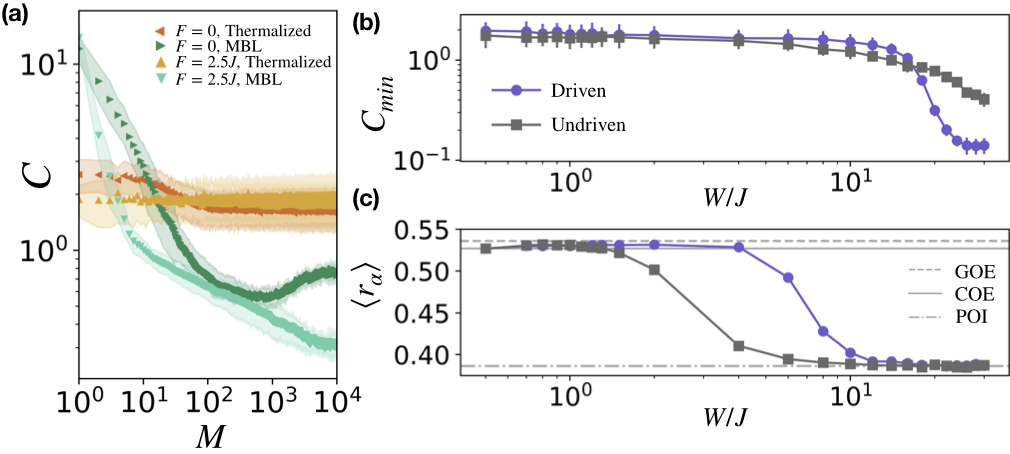}
\caption{\textbf{Training analog quantum systems in the Hilbert space:} (a) The lowest cost function at each training step $M$ for $F=0$ and $F=2.5J$. The thermalized and and the MBL phases are obtained with $W=1J$ and $W=20J$,  respectively. The shaded areas represent standard deviations.  (b) The cost function at $M=10^4$ as a function of $W$. The results are averaged over 10 dataset, i.e., 10 realizations of $\{a_i,b_i\}$ in Eq. (\ref{eq:bm}). Each dataset consists of $3000$ samples. (c) The averaged level spacing $\langle r_{\alpha}\rangle$ at $M=10^4$ as a function of $W$. ($L=9$, $\omega=8J$, $h=2.5J$,$k_BT_0=J$ and $D=200$.)}
\label{fig5}
\end{figure*}

\subsubsection{Training results}

The training results are shown in Fig. \ref{fig5}(a). As expected, the system in the thermalized phase cannot be trained. The cost function from the thermalized case with $F=2.5J$ saturates at $C\sim 2$ already at the first layer. In the $F=0$ case, the cost function starts at around $C\sim 3.5$ and then falls down to saturate at $C\sim 2$, the same value as the driven case, when $M\sim 50$. For the MBL case with $F=0$, during $M \lesssim 10^2$, the cost function steadily decays to $0.7$ . Then during $10^2 \lesssim M \lesssim 10^3$, the cost function continues to decay with a slower rate. Interestingly, after $M\gtrsim 10^3$, the cost function increases and saturates at $0.7$ when $M \sim 10^4$. In contrast, for the MBL case with $F=2.5J$, the cost function goes down steadily when $M\lesssim 10$. Then, the cost function further decays monotonically with a slower rate to saturate at $C\sim 0.1$ at $M\sim 10^4$. This results show that the learning behavior changes qualitatively depending on the phase and the timescale of the system. The best learning accuracy is obtained with the MBL phase with $F=2.5J$.

In Fig. \ref{fig5}(b), we plot the final learning results as a function of $W$ for $F=0$ and $F=2.5J$. For comparison, in Fig. \ref{fig5}(c), we also plot the averaged level spacing $\langle r_{\alpha}\rangle$ as a function of $W$ for both cases. In the $F=2.5J$ case, the final learning accuracy shows a transition between the trainable and the untrainable regimes, which corresponds roughly to the phase transition between the CUE and the POI statistics. In the $F=0$ case, the system moves towards the trainability regime as $W$ approaches $30J$. However, we stop our calculation here as the training takes too long to converge when $W>30J$ \footnote{For $M=10^4$, it takes approximately $1,200$ CPU hours to generate Fig. \ref{fig5}(b) using Intel Xeon E5-2690 v3 processors.}. Nevertheless, our present results are sufficient to conclude that the drive leads to a better learning accuracy for this timescale. We conjecture that, once the system in the undriven case fully reaches the trainable regime, the learning accuracy should monotonically decreases with $M$ until saturation.

\subsubsection{Temporal correlations enabled by MBL}

To understand different learning accuracy in different phases, we calculate the KL divergence between  $p(\underline{z};\underline{\Theta}_M)$ and $p(\underline{z};\underline{\Theta}_{M+\delta m})$ to measure the temporal correlations or the `memory' between outputs at different layers. In Fig. \ref{fig3}(d), we plot such KL divergence as a function of $\delta m$, averaged over various $M$'s. In the thermalized phase, we find that there are no temporal correlations between layers. This is expected as each layer has chaotic dynamics which is highly sensitive to any small changes introduced to the system. In contrast, in the MBL phase, the system displays short-term memory that decays with $\delta m$. The MBL dynamics with $f(t)=0$ has the longest memory. This memory were exploited during the training to improve trainability of the system.

\section{Conclusions}

In this work, we have throughly analyzed the expressibility and trainability of parameterized analog quantum many-body systems. We show that both thermalized and MBL dynamics with and without the modulation $f(t)$ are capable of reaching the quantum supremacy regime, indicating high expressibility beyond any classical models. In the context of generative modeling, we show that chaoticity prevents systematic optimization of the system. However, the latter can be qualitatively improved by the MBL dynamics. In the future, it would be interesting to analyze scalability and generalizability of our models as well as more complex training protocol for efficient optimization.  

\section{Acknowledgement}
This research is supported by the National Research Foundation, Prime Minister's Office, Singapore and the Ministry of Education, Singapore under the Research Centres of Excellence programme. It was also partially funded by Polisimulator project co-financed by Greece and the EU Regional Development Fund, the European Research Council under the European Union's Seventh Framework Programme(FP7/2007-2013)/ Ninnat Dangniam is supported by the National Natural Science Foundation
of China (Grant No. 11875110).

\bibliography{ref}

\end{document}